\documentclass[letterpaper]{article} %
\usepackage[preprint]{aaai2027}  %
\usepackage[hyphens]{url}  %
\usepackage{graphicx} %
\usepackage{natbib}  %
\usepackage{caption} %
\usepackage{amsmath,amsfonts}
\usepackage{algorithm}
\usepackage{algorithmic}
\usepackage{enumitem}
\usepackage{siunitx}

\def\CC{{\mathbb C}}
\def\RR{{\mathbb R}}
\usepackage{circledsteps}

\usepackage{amssymb}%
\usepackage{pifont}%
\newcommand{\cmark}{\ding{51}}%
\newcommand{\xmark}{\ding{55}}%
\usepackage{newfloat}
\usepackage{listings}
\DeclareCaptionStyle{ruled}{labelfont=normalfont,labelsep=colon,strut=off} %
\floatstyle{ruled}
\newfloat{listing}{tb}{lst}{}
\floatname{listing}{Listing}

\usepackage{booktabs}
\usepackage{multirow}
\nocopyright

\title{REDnet: Recursive Encoder and Decoder for Speech Separation under Unknown Number of Speakers and Variable Number of Microphones}
\author{
    Fulin Wu and Zhong-Qiu Wang
}
\affiliations{

    Department of Computer Science and Engineering, Southern University of Science and Technology, Shenzhen, China\\
    12532594@mail.sustech.edu.cn, wang.zhongqiu41@gmail.com
}

\begin{document}

\maketitle

\begin{abstract}
We propose \textit{recursive encoder and decoder} (RED) for building a single deep neural network (DNN) model that can separate multi-speaker mixtures containing unknown numbers of speakers and variable numbers of microphones arranged in an unknown geometry, a task that has not been studied yet.
The decoder of RED recursively detects whether there are active speakers left and separates one speaker at a time.
It is designed to be trained in an end-to-end fashion to improve separation performance.
The encoder of RED recursively encodes each microphone channel of the input mixture, sequentially incorporating spatial cues.
Combining both, the DNN can be trained to separate mixtures not only with unknown numbers of speakers but also with variable numbers of microphones, achieving state-of-the-art performance on multiple public datasets.
\end{abstract}

\section{Introduction}

In many artificial intelligence applications, the sensors inevitably record a mixture of multiple concurrent signals.
It is often necessary to unmix them before further processing.
One prominent example, in audio signal processing, is speech separation (\textit{a.k.a.}, the cocktail party problem) \cite{McDermott2009,WDLreview,li2023pgss,fan2025bsdb,xu2025tiger,xu2025arraydps,Li2026sepprune}, where multiple speakers often talk concurrently in natural conversations, and their mixed speech needs to be separated from each other before further processing.
In realistic scenarios, the number of concurrent speakers is often unknown and needs to be estimated, and, in addition, the recording device can be a single microphone or an array of microphones arranged in a geometry not observed during training.
This motivates us to build a single, general-purpose model that can flexibly separate
multi-speaker mixtures across a broad range of speaker counts, microphone counts, and microphone geometries.
This requires the model to solve two key challenges together: producing outputs for an unknown number of speakers based on inputs with a diverse number of microphones arranged in various and possibly unknown geometries.
This is particularly challenging for deep neural networks (DNN) based models as
the input and output dimensions are usually fixed.

There have been studies attacking the two challenges separately but not jointly.
For example, DNN modules such as TAC \cite{Luo2020e2e} and VarArray \cite{Yoshioka2022VarArray} that iteratively combine channel-specific features with globally aggregated cues across all microphones have been proposed to separate mixtures
recorded by variable numbers of microphones.
However, the number of speakers to estimate is assumed known and fixed, which limits their application scenarios.
On the other hand, SepTDA \cite{lee2024boosting} and SR-CorrNet
\cite{Shin2026SRCorrNet} propose a transformer-decoder-based attractor module to deal with unknown numbers of speakers, but they assume that the number of input microphones and their geometry are fixed between training and testing.
Recently, FlexIO \cite{masuyama2026flexio} trains a single DNN to handle speech separation and enhancement under known numbers of speakers and variable numbers of microphones, but it assumes that the number of sources and the source types are known a priori, which restricts its practical deployment.
In summary, there still lacks a unified solution that can address both challenges at the same time.

In this context, we make the first attempt towards solving both challenges inside a single DNN.
We propose \textit{recursive encoder and decoder network} (REDnet), equipped with a recursive encoder (RE) and a recursive decoder (RD).
RE recursively encodes each microphone, sequentially exploiting spectral and spatial cues for separation.
RD, building upon a monaural, recursive speech separation paradigm named OR-PIT \cite{takahashi2019recursive}, recursively performs speaker detection and separates one speaker at a time until no speakers are detected.
We summarize our contributions as follows:
\begin{itemize}[leftmargin=*,noitemsep,topsep=0pt]
\item We are the first to train a single DNN to perform speech separation under unknown numbers of speakers and variable numbers of microphones arranged in unknown geometries.
\item We propose RE, a new mechanism that recursively encodes information of each microphone.
We observe that RE captures spatial cues better than TAC.
\item We propose RD, which, building upon the recursive separation paradigm in OR-PIT \cite{takahashi2019recursive}, further improves the performance via end-to-end training, refined loss functions, and speaker interaction modules.
\item The proposed shared recursive design allows the encoder and decoder to be readily integrated into a unified framework (e.g., REDnet), which can aggregate spectral and spatial information across variable numbers of microphones to separate unknown numbers of speakers.
\end{itemize}
\noindent Evaluation results on multiple public datasets
show that REDnet obtains state-of-the-art performance, exhibiting strong generalizability to real-recorded signals.

\section{Related Work}

\subsection{Related work with RD}

To separate unknown numbers of speakers, there are roughly four approaches.
\Circled{\footnotesize 1}
\textit{Separate-then-select methods} train one model for each speaker count or train a single model to produce a fixed, maximum number of estimates, and subsequently identify valid ones \cite{Nachmani2020gatedDPRNN,luo2020separating,kim2023investigation,kim2023training}.
However, they require additional post-selection and always perform separation up to the assumed maximum number of speakers, regardless of the actual speaker count.
\Circled{\footnotesize 2}
\textit{Count-aware methods} jointly count speakers and route a shared representation to speaker-count-specific separation heads \cite{chazan2021single,zhu2021multi}.
Their performance relies on counting accuracy, and they have to pre-define speaker counts the model can handle.
\Circled{\footnotesize 3}
\textit{Attractor-based methods} estimate a variable number of speaker embeddings and use them to separate speakers\cite{horiguchi2020end,chetupalli2022sepeda,Maiti2023eendss,chetupalli2023speaker,lee2024boosting,Shin2026SRCorrNet}.
Although they can accommodate unknown numbers of speakers, speaker selection occurs in a latent attractor space, making its relation to the final speech estimates less transparent.
\Circled{\footnotesize 4}
\textit{Recursive methods}
extract one speaker at each iteration, and stop when no
speaker is detected \cite{shi2018listen,kinoshita2018listening,takahashi2019recursive,shi2020sequence,jin2022coarse,alizadeh2025resepnet}.
However, existing recursive approaches generally rely on relatively limited separation backbones, which restricts their overall performance.

RD follows the recursive paradigm, but rather than
counting speakers independently or generating all latent attractors before separation, it tightly couples speaker separation with counting: the stopping decision is made directly from the current separation state, so the number of speakers is counted jointly with the recursive separation process.
Compared with separate-then-select methods, RD
is trained end-to-end by jointly performing separation and
counting, avoiding post-selection. 
Compared with prior recursive methods, RD employs stronger time-frequency-domain models, speaker interaction modules and refined training objectives, and is trained in an end-to-end fashion, 
thereby realizing better separation.

\subsection{Related work with RE}
\label{sec:variable_microphones}
There are roughly three
approaches for handling variable numbers of microphones.
\Circled{\footnotesize 1}
\textit{Beamforming methods} \cite{Heymann2016smeab,Erodgan2016MVDR,Luo2019FasNet} estimate
linear filters to combine microphone signals, naturally accommodating any number of microphones.
However, linear filtering alone often cannot realize sufficient separation especially when the number of microphones is limited and reverberation is strong.
Applying a monaural DNN for post-filtering cannot fully exploit spatial cues \cite{Wang2018iCombineSpectralSpatial}.
\Circled{\footnotesize 2}
\textit{Average-based aggregation methods}, such as TAC,
extract features from individual microphones and combine each feature with a global representation obtained by averaging across microphones \cite{Luo2020e2e,Yoshioka2022VarArray,Zhang2023Diverse,kim2024enhanced,masuyama2026flexio}.
This design
can process a variable number of microphones without assuming a fixed geometry.
However, averaging-based aggregation compresses all microphone signals into a single global feature and may weaken channel-specific cues and fine-grained inter-channel patterns.
Repeated TAC operations, which are necessary to make TAC work, dramatically increase computation.
\Circled{\footnotesize 3}
\textit{Explicit inter-channel modeling methods} model 
multiple microphones using 
cross-attention \cite{wang2020neural,wang2021continuous,Pandey2022TADRN,wang23SDNet,guo2024graph,guo2025co,masuyama2026flexio}.
These methods enable adaptive information exchange among microphones and can accommodate diverse microphone counts and array geometries.
However, explicitly modeling inter-channel dependencies generally incurs architectural complexity and computational overhead.

Our proposed RE progressively integrates one microphone at each recursion step by jointly modeling the current microphone signal and the information accumulated from previously processed microphone signals.
This avoids both the information compression introduced by TAC-style global averaging and the quadratic cost associated with explicit pairwise inter-channel modeling.
By interacting only the current microphone with the accumulated representation, RE preserves channel-specific information, while maintaining linear complexity with respect to the number of microphones.

\subsection{Related work with REDnet}

There are recent efforts developing a single DNN model for flexible speech separation and enhancement, such as FlexIO \cite{masuyama2026flexio}.
However, its flexibility relies on externally provided speaker prompts and does not require the model to infer the number of speakers.
Our study takes a further step towards a universal speech separation and enhancement system by jointly dealing with variable numbers of microphones and unknown numbers of speakers.

\begin{figure*}[t]
\centering
\includegraphics[width=1.0\textwidth]{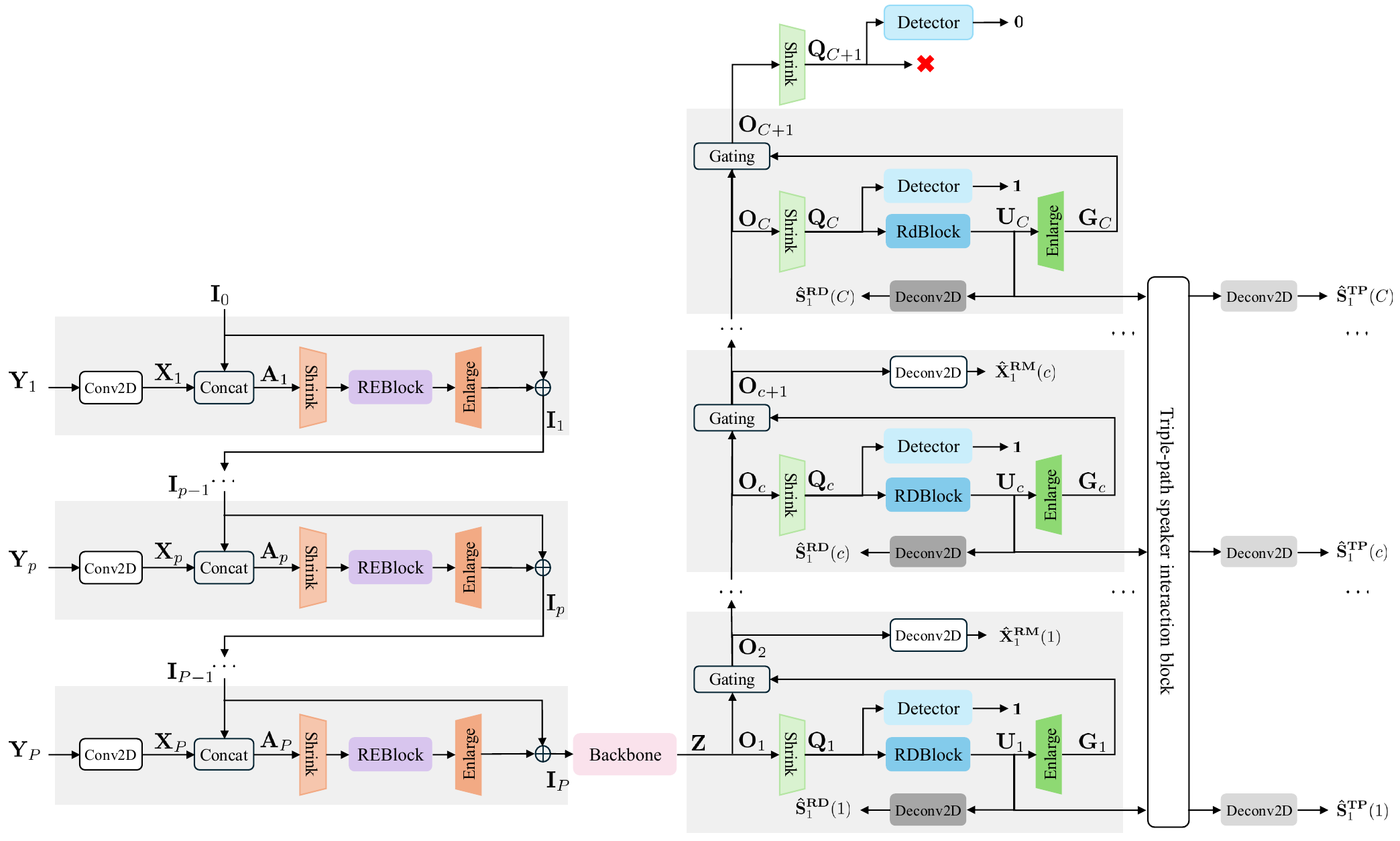} %
\caption{
Overview of proposed REDnet (best viewed in color). Blocks with the same name and color share the same parameters.
See Fig. \ref{block_figure_label} for an illustration of REBlock, RDBlock, Backbone, triple-path speaker interaction block, and speaker detector.
}
\label{fig:RED}
\end{figure*}

\section{Problem Formulation}

In a noisy-reverberant environment containing $C$ concurrent speakers, the signal recorded by a microphone array with $P$ microphones can be formulated, in the short-time Fourier transform (STFT) domain, as follows:
\begin{align}\label{pyysical_model_TF}
Y_p(t,f) &= \sum\nolimits_{c=1}^C S_p(c,t,f) + V_p(t,f) \in \CC,
\end{align}
where $t$, $f$, $p$ and $c$ respectively index $T$ time frames, $F$ frequency bins, $P$ microphones and $C$ speakers.
At microphone $p$, $Y_p$, $S_p(c)$ and $V_p$ respectively denote the spectrograms of the mixture, target-speech signal of speaker $c$, and non-target signals.
Without loss of generality, we designate the first microphone as the reference microphone, and our goal is to estimate the target signals $\{S_1(c)\}_{c=1}^{C}$ at the reference microphone $1$ based on the observed mixtures $\{Y_p\}_{p=1}^{P}$.
Notice that, in realistic applications, different mixtures often contain different numbers of speakers and are recorded by different microphone arrays with different numbers of microphones.
That is, $P$ and $C$ can vary from mixture to mixture.
Our objective is to train a single DNN model that can perform accurate speech separation for a wide range of speaker and microphone counts.
Notice that, at run time, the number of microphones is readily available from the input mixture, whereas the number of active speakers needs to be estimated.

\section{REDnet}\label{sec:red}

Fig. \ref{fig:RED} illustrates our proposed REDnet, which consists of an RE module to deal with variable numbers of microphones, a Backbone to model spectral, temporal, and spatial patterns, and an RD module for dealing with unknown numbers of speakers.
This section describes each of them.

\subsection{Recursive encoder}

Given an STFT-domain representation of the mixture signal at each microphone $p$, $\mathbf{Y}_p\in \mathbb{R}^{2 \times T \times F}$, where the first dimension stacks the real and imaginary (RI) components, a two-dimensional convolution (Conv2D) module is first applied to embed each microphone signal independently, obtaining an embedding $\mathbf{X}_p \in \mathbb{R}^{D \times T \times F}$.
Next, starting from the reference microphone $1$, RE recursively processes the $P$ embeddings, progressively aggregating their information into a unified representation.
Let $\mathbf{I}_{p} \in \mathbb{R}^{D \times T \times F}$ denote the output
after the $p$-th recursion, which represents the accumulated feature after exploiting information from the first $p$ microphones.
The recursive representation starts froms $\mathbf{I}_{0} = \mathbf{0}$.
At the $p$-th recursion, $\mathbf{I}_{p-1}$
is concatenated with the representation of the current microphone signal $\mathbf{X}_{p}\in \mathbb{R}^{D \times T \times F}$ along the first dimension: $\mathbf{A}_{p} = \operatorname{concat} \left( \mathbf{I}_{p-1}, \mathbf{X}_{p} \right) \in \mathbb{R}^{2D \times T \times F}$,
which is then projected by a shrink layer based on point-wise Conv2D to $E \times T \times F$ and subsequently modeled by an REBlock.
Next, an enlarge layer, also based on point-wise Conv2D, restores the feature dimension to $D$. Its output is added to the previous accumulated representation $\mathbf{I}_{p-1}$ through a residual connection, yielding
$\mathbf{I}_{p} \in \mathbb{R}^{D \times T \times F}$, which aggregates information from the first $p$ microphones.
After all the $P$ microphones have been processed, the final aggregated representation $\mathbf{I}_{P}$ is fed to a Backbone module to further refine the representation, obtaining $\mathbf{Z} \in \mathbb{R}^{D \times T \times F}$ for the recursive decoder to separate an unknown number of speakers.
The details of the REBlock and Backbone modules will be described later in Section \ref{DNN_config}.

\subsection{Recursive decoder}

Given the Backbone output $\mathbf{Z} \in \mathbb{R}^{D \times T \times F}$, RD recursively extracts one speaker at each iteration
until no speaker is detected.
We set $\mathbf{O}_{1} =\mathbf{Z}$ to start the decoding process.
At the $c$-th iteration, the residual representation $\mathbf{O}_c \in \mathbb{R}^{D \times T \times F}$ is projected by a shrink layer (based on point-wise Conv2D) to $E \times T \times F$.
Based on the shrunk feature, a speaker detector (which is an MLP to be described in Section \ref{DNN_config}) first determines whether there are active speakers.
If no speaker is detected, the recursive decoding process terminates; otherwise, the shrunk feature is passed to an RDBlock to obtain a feature representation of the $c$-th speaker, denoted as $U_c\in \mathbb{R}^{D \times T \times F}$, which can be fed to a 2D deconvolution (Deconv2D) layer to produce an estimate of the RI components of speaker $c$,
$\widehat{\mathbf{S}}_1^{\text{RD}}(c) \in \mathbb{R}^{2 \times T \times F}$, on which a loss function can be defined to train the network.

Meanwhile, an enlarge layer based on point-wise Conv2D is applied to $U_c$ to restore the feature dimension from $E$ to $D$, producing $\mathbf{G}_{c} \in \mathbb{R}^{D \times T \times F}$, which is used to derive a gating mechanism for updating the residual representation:
\begin{equation}
\mathbf{O}_{c+1} = \mathbf{O}_{c} \odot \big(1 - \operatorname{sigmoid}(\mathbf{G}_{c}) \big).
\end{equation}
This gating mechanism mimics
time-frequency (T-F) masking in speech separation and enhancement \cite{WDLreview}, since $\mathbf{G}_{c}$, derived from $\mathbf{U}_{c}$ which is regularized to reconstruct speaker $c$, likely only contains information about speaker $c$, and hence the computed gate, $1 - \operatorname{sigmoid}(\mathbf{G}_{c})$, could \textit{peel off} speaker $c$ from the current residual representation $\mathbf{O}_c$, yielding a new residual representation $\mathbf{O}_{c+1}$ that potentially only contains information about the speakers that are not yet separated.
This mechanism could be useful for recursive speech separation, as the embeddings in later iterations should not contain information about the speakers that have already been separated in earlier iterations.

This also motivates us to design a loss to penalize each $\mathbf{O}_{c+1}$ to encourage them to contain information only about the remaining speakers not yet separated.
We realize this by applying a Deconv2D layer to $\mathbf{O}_{c+1}$ to compute a spectrogram (denoted as $\widehat{\mathbf{X}}_1^{\text{RM}}(c+1) \in \mathbb{R}^{2 \times T \times F}$, with ``RM'' meaning ``remaining mixture''), which is then penalized by the summation of the target signals of the remaining speakers (i.e., $\sum_{c'=c+1}^{C}S_q(c')$).
We skip this for $\mathbf{O}_{C+1}$, as it is expected to have no speakers left in it.

On the other hand, since the speakers are serially estimated in different recursion steps and hence lack direct interaction, a triple-path speaker interaction module is attached to jointly model the speaker embeddings produced by RDBlock across all the recursion steps.
This could allow information to be exchanged when the spectrograms of different speakers are estimated.
In detail, we feed $\{U_c\}_{c=1}^C$ to the speaker interaction module to refine each speaker embedding, and then use a shared Deconv2D layer to estimate the spectrograms of each speaker.
For each speaker $c$, we denote the estimate as $\mathbf{S}_1^\text{TP}(c)\in \RR^{2\times T\times F}$, where ``TP'' means ``triple-path''.

\textbf{Training setup}.
During training, the oracle number of speakers, $C$, is used to unroll the RD module for $C$ iterations.
An additional speaker-detection step is performed to have the speaker detector learn to stop the recursion.

\textbf{Inference setup}.
At run time, the number of speakers needs to be estimated.
The recursive decoding continues until the predicted probability of the speaker detector falls below a pre-defined threshold $\tau$ (set to $0.5$).
This way, an estimated number of speakers, $\hat{C}$, can be used for separation.

\begin{figure}[t]
\centering
\includegraphics[width=0.5\textwidth]{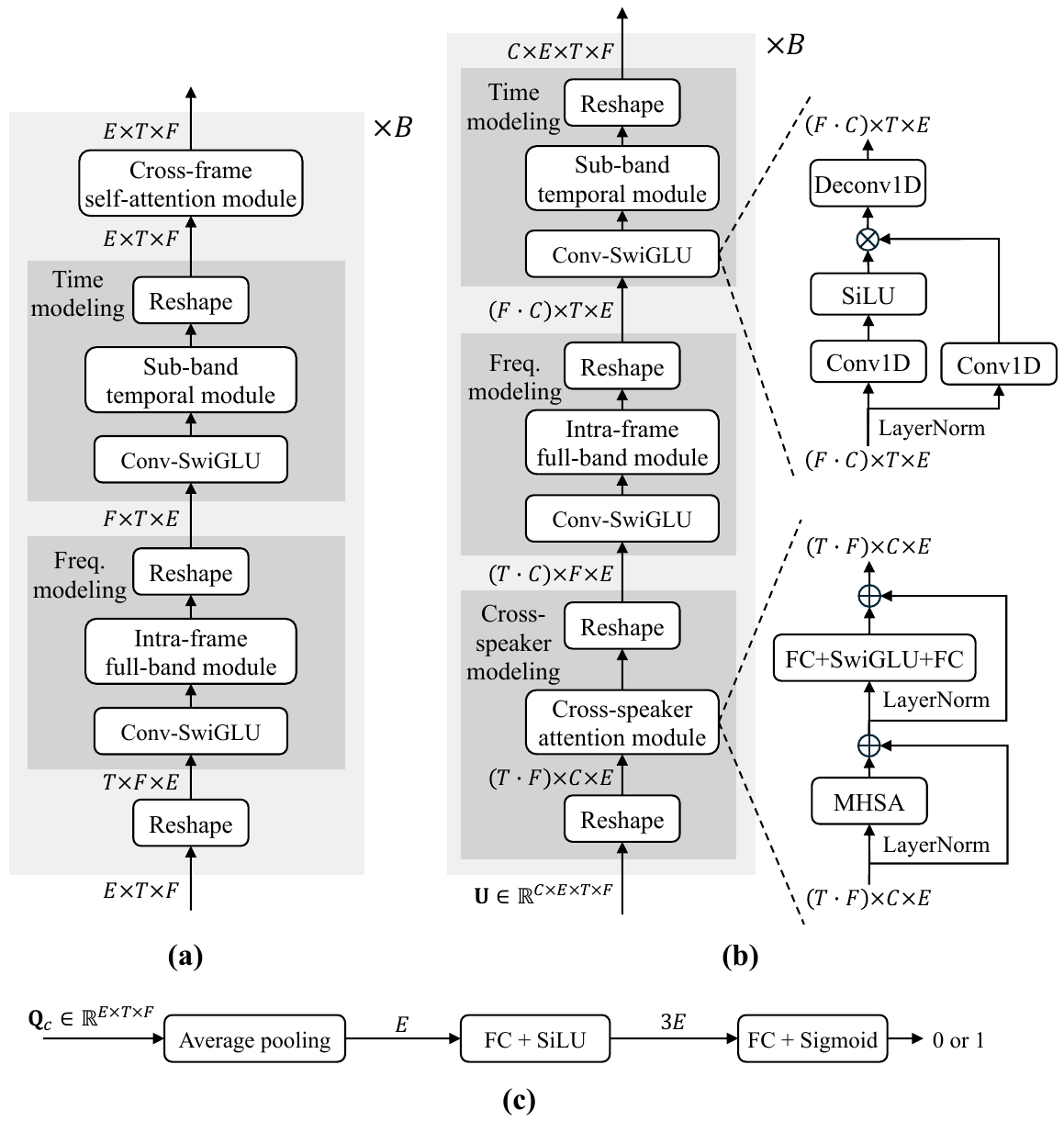} %
\caption{
Illustration of (a) REBlock, RDBlock and Backbone;
(b) speaker interaction block;
and (c) speaker detector.
}
\label{block_figure_label}
\end{figure}

\subsection{
DNN architectures of various blocks
}\label{DNN_config}

Fig. \ref{block_figure_label}(a) shows the architecture of REBlock, Backbone, and RDBlock.
It is based on the widely-adopted TF-GridNet block \cite{Wang2023TFGridNet}, which stacks an intra-frame full-band module, a sub-band temporal module, and a cross-frame self-attention module for efficient time and frequency modeling.
We make two minor modifications: replace all PReLU activations with SiLU \cite{ELFWING2018silu}, and insert a Conv-SwiGLU module proposed in TF-LocoFormer \cite{Saijo2024LocoFormer} before the intra-frame full-band module and sub-band temporal module to better model local patterns.

The triple-path speaker interaction block in Fig. \ref{block_figure_label}(b) sequentially performs cross-speaker, frequency and time modeling, where the latter two share the same architecture as those in Fig. \ref{block_figure_label}(a).
For cross-speaker modeling, we reshape the input tensor $C\times E\times T\times F$ to $(T\cdot F)\times C\times E$,
view it as $T\cdot F$ sequences each with $C$ steps, and use a multi-head self-attention (MHSA) layer followed by a one-hidden-layer MLP with SwiGLU activation to model speaker interaction.

The speaker detector in Fig. \ref{block_figure_label}(c) takes an $E\times T\times F$ tensor as input, and applies average pooling over time and frequency, obtaining an $E$-dimensional vector, which is then processed by a one-hidden-layer MLP with a SiLU hidden activation, followed by a sigmoid activation to estimate the probability of whether at least one speaker is active.

\subsection{Loss functions}
\label{sec:loss}

The scale-invariant signal-to-noise ratio (SI-SNR) loss \cite{LeRoux2019} is leveraged to penalize estimated signals.
Let $w_{\mathrm{RD}}$, $w_{\mathrm{RM}}$, $w_{\mathrm{TP}}$, and $w_{\mathrm{BCE}}$ denote weights for different loss terms; the overall loss function is defined as follows:
\begin{align}
\mathcal{L}
= w_{\mathrm{RD}} \cdot \mathcal{L}_{\mathrm{RD}} +w_{\mathrm{RM}} \cdot \mathcal{L}_{\mathrm{RM}} 
+w_{\mathrm{TP}} \cdot \mathcal{L}_{\mathrm{TP}} +w_{\mathrm{BCE}} \cdot \mathcal{L}_{\mathrm{BCE}}. \nonumber
\end{align}
The first three terms penalize $\widehat{\mathbf{S}}_1^{\text{RD}}$, $\widehat{\mathbf{S}}_1^{\text{RM}}$ and $\widehat{\mathbf{S}}_1^{\text{TP}}$, respectively.
The fourth term, $\mathcal{L}_{\mathrm{BCE}}$, is the binary cross-entropy (BCE) loss for training the speaker detector.
For a mixture containing $C$ speakers, the label
sequence for training is $[1,\dots,1,0]$, consisting of $C$ ones followed by a zero, where $1$ indicates that at least one speaker exists and $0$ means no speakers exist.

To successfully train REDnet for separation, we need to resolve the label-permutation problem before loss computation.
If $w_{\mathrm{TP}}$ is set positive, we identify the best permutation by using permutation invariant training \cite{Kolbak2017} on $\mathcal{L}_{\mathrm{TP}}$, and use the best permutation to compute $\mathcal{L}_{\mathrm{RD}}$ and $\mathcal{L}_{\mathrm{RM}}$.
If $w_{\mathrm{TP}}$ is set to zero (i.e., the speaker interaction module is not used), we resolve the permutation based on $\mathcal{L}_{\mathrm{RD}}$, and apply the same permutation to compute $\mathcal{L}_{\mathrm{RM}}$.

\section{Experimental Setup}

Based on public datasets (see Table \ref{dataset_info}), we first separately validate
whether RE can handle variable numbers of microphones and whether RD can separate unknown numbers of speakers, followed by their integration (i.e., RED) to check whether both can be realized simultaneously.
As different datasets often have different definitions of target speech, following their design, the target speech can be reverberant (denoted as ``r''), direct-path (``d'') and direct-path plus $50$ ms of early reflections (``d+e'').
This section describes the datasets, baselines, system configurations, and evaluation metrics.

\subsection{Datasets for REnet, RDnet and REDnet}

For RD, we construct RDnet by replacing the recursive encoder in REDnet with a Conv2D layer which embeds input mixtures recorded by a fixed number of microphones from $2P \times T \times F$ to $D \times T \times F$ (where $P$ is fixed).
We train and test RDnet on the \textit{WSJ0-\{2,3,4,5\}mix} dataset \cite{Nachmani2020gatedDPRNN}, which is monaural, anechoic, noise-free, and consists of mixtures with $2$ to $5$ speakers.
In addition, to show the effectiveness of RDnet in noisy-reverberant conditions,
we train and test it based on a noisy-reverberant version of WSJ0-\{2,3,4,5\}mix, named \textit{WSJ0-\{1,2,3,4,5\}mix-nr} \cite{saijo2023single}, which
additionally provides single-speaker signals to evaluate models' effectiveness at speech enhancement.

For RE, we construct REnet by replacing the recursive decoder in REDnet with a Deconv2D layer to predict the RI components of target signals, assuming a fixed number of speakers to separate (i.e., from $D \times T \times F$ to $2C \times T \times F$, with $C$ fixed).
We perform training and testing on a dataset introduced in the TAC paper \cite{Luo2020e2e}.
The dataset is designed for multi-channel two-speaker separation in noisy-reverberant conditions, covering both ad-hoc- and fixed-geometry array setups (denoted respectively as \textit{TAC-ad-hoc} and \textit{TAC-fixed-array}).
The models are trained only on the ad-hoc array data, with the fixed-geometry array data reserved for evaluating generalizability to unseen array geometries.

For RED, we strictly follow the experimental setup of FlexIO \cite{masuyama2026flexio}, using the same data to train and test REDnet so that a direct performance comparison can be made.
The FlexIO setup combines five datasets in speech separation and enhancement, including \textit{WSJ0-\{2,3\}mix} \cite{Hershey2016}, \textit{WHAM!} \cite{Wichern2019}, anechoic and reverberant \textit{WHAMR!} (denoted respectively as WHAMR!-A and WHAMR!-R) \cite{Maciejewski2020}
\textit{WSJ1-CHiME4} \cite{Scheibler2021} and \textit{CHiME-4} \cite{Barker2017CHiME3}, which have different numbers of speakers and microphones as well as acoustic conditions.
This setting is well-suited to our task, as REDnet is designed to separate unknown numbers of speakers given variable numbers of microphones.

\begin{table}[t]
    \centering
    \scriptsize
    \sisetup{table-format=2.1,round-mode=places,round-precision=1,table-number-alignment=center,detect-weight=true,detect-inline-weight=math}
    \setlength{\tabcolsep}{1.25pt}
    \begin{tabular}
        {
        l
        c %
        c %
        c %
        c %
        c %
        }
        \toprule
        Model & Dataset & Ver. / SR (kHz)  & $P$ & $C$ & $S_1$ \\
        \midrule
        \multirow{2}{*}{REnet} & TAC-ad-hoc & -- / 16 & \{2,3,4,5,6\} & \{2\} & \multirow{2}{*}{r} \\
         & TAC-fixed-array & -- / 16 & \{\}, \{6\} & \{\}, \{2\} &  \\
        \midrule
        RDnet1 & WSJ0-\{2,3,4,5\}mix & min / 8 & \{1\} & \{2,3,4,5\} & d \\
        RDnet2 & WSJ0-\{1,2,3,4,5\}mix-NR & min / 8 & \{1\} & \{1,2,3,4,5\} & d+e \\
        \midrule
        \multirow{5}{*}{REDnet} & WSJ0-\{2,3\}mix & min / 16 & \{1\} & \{2,3\} & \multirow{5}{*}{d} \\
         & WHAM! & min / 16 & \{1\} & \{1,2\} & \\
         & WHAMR!-\{A,R\} & min / 16 & \{1,2\} & \{1,2\} & \\
         & WSJ1-CHiME4 & -- / 16 & \{2,4\}, \{3\} & \{2,3\} & \\
         & CHiME-4 & -- / 16 & \{1,2,4\}, \{3, 5\} & \{1\} & \\
        \bottomrule
    \end{tabular}
    \caption{
    Datasets, including version (Ver.) and sampling rate (SR), used for each model. REnet is trained on \textit{TAC-ad-hoc} and tested on \textit{TAC-ad-hoc} and \textit{TAC-fixed-array}.
    RDnet1 is trained and tested on \textit{WSJ0-\{2,3,4,5\}mix}, and RDnet2
    on \textit{WSJ0-\{1,2,3,4,5\}mix-NR}. REDnet is trained and tested on the five datasets combined.
    In column ``$P$'', some rows have two sets of values separated by a comma.
    This means that the microphone counts in the right set are reserved for testing to check generalizability, not used for training.
    }
    \label{dataset_info}
\end{table}

\subsection{Miscellaneous system configurations}

REBlock, Backbone, and RDBlock share the same modified TF-GridNet architecture but have different hyper-parameters and model weights.
Using the notations in the TF-GridNet paper \cite{Wang2023TFGridNet}, for REBlock and RDBlock, we set $D=64$, $B=4$, $I=4$, $J=4$, $H=256$ and $L=4$; 
and for Backbone, we change $D=128$, $I=1$ and $J=1$.
For each Conv1D and Deconv1D in Conv-SwiGLU, the kernel size is $3$ and the stride is $1$.
In the cross-speaker attention module of the speaker interaction block, the MHSA layer uses $4$ attention heads and the expansion ratio in the MLP layer is set to $3$.
We stack $B=2$ speaker interaction blocks.

Each microphone signal is normalized to unit sample variance before any processing.
For STFT, the window size is $16$ ms, hop size $8$ ms, and the square-root Hann window is used as the analysis window.
The models are trained using the Adam optimizer.
The learning rate starts from $5\times10^{-4}$ and is halved if the validation loss is not improved in $3$ epochs.
The batch size is $1$. The segment size is $8$ seconds for training RDnet and REDnet, and $4$ seconds for REnet.
Gradient clipping with a maximum $L_2$ norm of $1.0$ is applied to stabilize training.
Mixed-precision with bfloat$16$ is used in training.

The evaluation metrics include SI-SDR \cite{LeRoux2019}, BSSEval SDR \cite{vincent2006bssevalsdr}, signal-to-interference ratio (SIR) \cite{vincent2006bssevalsdr}, narrow-band PESQ (NB-PESQ) \cite{rix2001NBPESQ} and wide-band PESQ (WB-PESQ) \cite{ITU2007WBPESQ}, STOI \cite{taal2011STOI}, and DNSMOS (OVRL) \cite{reddy2022dnsmos}.
For SI-SDR,
on some datasets we report SI-SDRi, which is the SI-SDR improvement of estimated signals over unprocessed mixtures.

Following previous studies \cite{lee2024boosting,Shin2026SRCorrNet},
at run time, if the estimated number of speakers $\hat{C}$ exceeds $C$, an optimal subset of estimated signals is selected
for metric computation; and
if $\hat{C}<C$, the predictions are padded with $C - \hat{C}$
signals whose values are all $10^{-8}$.

\subsection{Baseline systems}

Since our models are trained on public datasets, we can readily compare the results of our models with existing ones.

For REnet, we consider TAC \cite{Luo2020e2e} as the baseline, as it is a representative model for speech separation under variable numbers of microphones.
To use a computation cost similar to RE, we implement TAC following the design in USES \cite{Zhang2023Diverse} and equip it with the same ``shrink--modeling--enlarge'' structure.
At each TAC layer, different microphone embeddings first go through a shrink layer to reduce the feature dimension, followed by a T-F modeling block that shares the same architecture and configuration as REBlock except for the number of blocks $B$, and they are then fused across microphones by the TAC operation, and go through an enlarge layer to restore the original feature dimension, followed by a residual connection.
REBlock uses $4$ modified TF-GridNet blocks, whereas TAC uses $1$ such block per TAC layer and stacks $4$ TAC layers to achieve a similar computation cost.

For RDnet, we compare its performance with previous best methods (see Table \ref{tab:wsj0_mix_sota}) on the \textit{WSJ0-\{2,3,4,5\}mix} dataset.
We further compare it with MUSE \cite{saijo2023single}, which proposes the \textit{WSJ0-\{1,2,3,4,5\}mix-NR} dataset.

For REDnet, we compare its performance mainly with FlexIO \cite{masuyama2026flexio}, which assumes oracle speaker count and source types at run time.

\begin{table}[t]
    \centering
    \scriptsize
    \sisetup{table-format=2.1,round-mode=places,round-precision=1,table-number-alignment=center,detect-weight=true,detect-inline-weight=math}
    \setlength{\tabcolsep}{2pt}
    \begin{tabular}
        {
        c
        c
        c
        S[table-format=1.1,round-precision=1] %
        S[table-format=1.1,round-precision=1] %
        S[table-format=1.1,round-precision=1] %
        S[table-format=1.1,round-precision=1] %
        S[table-format=2.1,round-precision=1]
        S[table-format=2.1,round-precision=1]
        S[table-format=2.1,round-precision=1]
        S[table-format=2.1,round-precision=1]
        }
        \toprule
        & & \multirow{2}{*}[-5.5pt]{\begin{tabular}[c]{@{}c@{}}Conv- \\ SwiGLU\end{tabular}} & & & & & \multicolumn{4}{c}{SI-SDRi (dB)} \\
        \cmidrule(lr){8-11}
        {Gating}
        & {SiLU}
        & 
        & {$w_{\mathrm{RM}}$}
        & {$w_{\mathrm{RD}}$}
        & {$w_{\mathrm{TP}}$}
        & {$w_{\mathrm{BCE}}$}
        & {2-mix}
        & {3-mix}
        & {4-mix}
        & {5-mix} \\
        \midrule
        \xmark & \xmark & \xmark & 0.0 & 1.0 & 0.0 & 1.0 & 22.852567145632953 & 21.99208777765423 & 19.349479395922057 & 17.56200452217619 \\
        \cmark & \xmark & \xmark & 0.0 & 1.0 & 0.0 & 1.0 & 23.21065235151388 & 22.323057607770796 & 19.68710077706104 & 17.81789446432609 \\
        \midrule
        \cmark & \cmark & \xmark & 0.0 & 1.0 & 0.0 & 1.0 & 23.331062957164274 & 22.380396677809667 & 19.74069331086291 & 17.973194897126174 \\
        \cmark & \cmark & \cmark & 0.0 & 1.0 & 0.0 & 1.0 &  23.98552983180098 & 23.16458814481108 & 20.572195241979017 & 18.840415294568405 \\
        \midrule
        \cmark & \cmark & \cmark & 0.0 & 0.0 & 1.0 & 1.0 & 24.57175114760282 & 24.18072170984864 & 21.88945473303859 & 20.27208408441812 \\
        \cmark & \cmark & \cmark & 0.0 & 0.1 & 1.0 & 1.0 & \bfseries 24.698563279181542 & 24.3602125687549 & 21.84146950223343 & 20.307312551007588 \\
        \cmark & \cmark & \cmark & 0.1 & 0.1 & 1.0 & 1.0 & \bfseries 24.71231843014044 & 24.39960800366978 & 22.165908619625366 & 20.64194578498928 \\
        \midrule
        \cmark & \cmark & \cmark & 0.1 & 0.1 & 1.0 & 2.5 & \bfseries 24.73534125783432 & \bfseries 24.521175449125625 & \bfseries 22.32029494830682 & \bfseries  20.877096186292444 \\
        \bottomrule
    \end{tabular}
    \caption{
    Ablation results of RDnet1 on WSJ0-\{2,3,4,5\}mix, assuming known speaker count at run time.
    }
    \label{tab:rd_ablation}

\vspace{0.3cm}

\centering
\scriptsize

\sisetup{
    table-format=2.1,
    round-mode=places,
    round-precision=1,
    table-number-alignment=left,
    table-text-alignment=left,
    detect-weight=true,
    detect-inline-weight=math
}
\setlength{\tabcolsep}{0.75pt}

\newcommand{\acc}[2]{\multicolumn{1}{l}{$#1_{(#2)}$}}
\newcommand{\na}{\multicolumn{1}{l}{--}}

\begin{tabular}{l *{4}{S}}
\toprule
 & \multicolumn{4}{c}{SI-SDRi (dB)} \\
\cmidrule(lr){2-5}
Model & {2-mix} & {3-mix} & {4-mix} & {5-mix} \\
\midrule

RSS \cite{takahashi2019recursive}             & \acc{14.8}{\text{N/A}} & \acc{12.6}{\text{N/A}} & \acc{10.2}{\text{N/A}} & \text{N/A} \\
gDPRNN \cite{Nachmani2020gatedDPRNN}       & \acc{18.6}{84.6} & \acc{14.6}{69.0} & \acc{11.5}{47.5} & \acc{10.4}{92.3} \\
SepEDA \cite{chetupalli2022sepeda}            & \acc{21.1}{99.8} & \acc{18.4}{97.0}
                         & \acc{14.4}{90.2} & \acc{11.6}{96.9} \\
ReSepNet \cite{alizadeh2025resepnet}           & \acc{21.2}{\text{N/A}} & \acc{19.2}{\text{N/A}}
                         & \acc{14.9}{\text{N/A}} & \acc{12.0}{\text{N/A}} \\
SepNet \cite{yang25SepNetEDCI}
                         & \acc{21.4}{99.9} & \acc{19.9}{99.5}
                         & \acc{16.9}{97.7} & \acc{14.3}{95.7} \\
SepTDA \cite{lee2024boosting}          & \acc{23.6}{99.9} & \acc{22.1}{95.9}
                         & \acc{19.5}{90.1} & \acc{16.9}{83.2} \\
SR-CorrNet-B \cite{Shin2026SRCorrNet}       & \acc{\mathbf{24.8}}{\mathbf{100.0}} & \acc{24.4}{\mathbf{99.7}}
                         & \acc{21.9}{97.7} & \acc{19.9}{96.9} \\
\midrule
RDnet1                 & \acc{24.7}{99.9} & \acc{\mathbf{24.5}}{\mathbf{99.7}}
                         & \acc{\mathbf{22.3}}{\mathbf{99.0}} & \acc{\mathbf{20.8}}{\mathbf{99.2}} \\
\bottomrule
\end{tabular}

\caption{
SI-SDRi comparison on WSJ0-\{2,3,4,5\}mix, all obtained based on estimated number of speakers.
Values in parentheses denote speaker-counting accuracy (\%).
}
\label{tab:wsj0_mix_sota}

\vspace{0.3cm}

\centering
\scriptsize

\sisetup{
    table-format=3.1,
    round-mode=places,
    round-precision=1,
    minimum-decimal-digits=1,
    table-number-alignment=center,
    detect-weight=true,
    detect-inline-weight=math
}

\setlength{\tabcolsep}{5pt}
\renewcommand{\arraystretch}{1.08}

\begin{tabular*}{\columnwidth}{
    @{\extracolsep{\fill}}
    l
    S[table-format=3.1,round-precision=1]
    S[table-format=3.1,round-precision=1]
    S[table-format=3.1,round-precision=1]
    S[table-format=3.1,round-precision=1]
    @{}
}
\toprule
Model
& \multicolumn{1}{c}{$C=2$}
& \multicolumn{1}{c}{$C=3$}
& \multicolumn{1}{c}{$C=4$}
& \multicolumn{1}{c}{$C=5$} \\
\midrule

SR-CorrNet-B
& 188.2
& 261.9
& 335.5
& 409.1 \\

RDnet1
& \bfseries 177.4
& \bfseries 222.5
& \bfseries 267.6
& \bfseries 312.7 \\

\bottomrule
\end{tabular*}

\caption{Computation cost in GMACs per second.
}
\label{tab:computational_cost}
\end{table}

\begin{table}[!]
\centering
\scriptsize
\sisetup{table-format=2.1,round-mode=places,round-precision=1,table-number-alignment=center,detect-weight=true,detect-inline-weight=math}
\setlength{\tabcolsep}{5pt}
\begin{tabular}{
    l
    S[table-format=2.1,round-precision=1]
    S[table-format=2.1,round-precision=1]
    S[table-format=2.1,round-precision=1]
    S[table-format=2.1,round-precision=1]
    S[table-format=2.1,round-precision=1]
}
\toprule
& \multicolumn{3}{c}{TAC-ad-hoc}
& \multicolumn{1}{c}{TAC-fixed-array} \\
\cmidrule(lr){2-4}
\cmidrule(lr){5-5}
Method
& {$P=2$}
& {$P=4$}
& {$P=6$}
& {$P=6$} \\
\midrule

TAC 
& 17.789 & 20.241 & 20.859 & 17.053 \\
REnet
& \bfseries 19.302 & \bfseries 22.254 & \bfseries 23.382 & \bfseries 18.518 \\

\bottomrule
\end{tabular}

\caption{SI-SDRi (dB) results on TAC-\{ac-hoc,fixed-array\}.
}
\label{tab:re-fixed}
\end{table}

\begin{table*}[t]
\centering
\scriptsize

\newcolumntype{D}{
  S[
    table-format=-2.1,
    round-mode=places,
    round-precision=1,
    table-number-alignment=center,
    detect-weight=true,
    detect-inline-weight=math
  ]
}

\newcolumntype{P}{
  S[
    table-format=1.2,
    round-mode=places,
    round-precision=2,
    table-number-alignment=center,
    detect-weight=true,
    detect-inline-weight=math
  ]
}
\sisetup{
    round-mode=places,
    round-precision=1,             
    table-format=2.1,              
    table-number-alignment=center, 
    detect-weight=true,             
    detect-inline-weight=math     
}
\setlength{\tabcolsep}{1.2pt}

\begin{tabular*}{\textwidth}{
  @{\extracolsep{\fill}}
  l
  *{10}{D}
  *{5}{P}
  @{}
}
\toprule

& \multicolumn{5}{c}{SI-SDR (dB)}
& \multicolumn{5}{c}{SDR (dB)}
& \multicolumn{5}{c}{NB-PESQ} \\
\cmidrule(lr){2-6}
\cmidrule(lr){7-11}
\cmidrule(lr){12-16}

Test set
& {Mixture} & {MUSE} & {RDnet2} & {MUSE$^{\star}$} & {RDnet2$^{\star}$}
& {Mixture} & {MUSE} & {RDnet2} & {MUSE$^{\star}$} & {RDnet2$^{\star}$}
& {Mixture} & {MUSE} & {RDnet2} & {MUSE$^{\star}$} & {RDnet2$^{\star}$} \\
\midrule

1-mix

&  4.519 & 11.5 & \bfseries 16.082 & 11.5 & \bfseries 16.082
&  6.718 & 13.9 & \bfseries 18.000 & 13.9 & \bfseries 18.000
& 2.094 & 3.01 & \bfseries 3.792 & 3.01 & \bfseries 3.792 \\

2-mix

& -1.622 &  8.4 & \bfseries 14.971 &  8.4 & \bfseries 14.936
& -0.783 & 10.1 & \bfseries 16.672 & 10.1 & \bfseries 16.655
& 1.594 & 2.45 & \bfseries 3.657 & 2.45 & \bfseries 3.656 \\

3-mix

& -4.577 &  5.4 & \bfseries 13.551 & 5.3 & \bfseries 13.532
& -3.737 &  7.0 & \bfseries 15.064 & 6.9 & \bfseries 15.055
& 1.393 & 2.05 & \bfseries 3.415 & 2.05 & \bfseries 3.414 \\

4-mix

& -6.414 &  2.5 & \bfseries 12.175 &  2.4 & \bfseries 12.165
& -5.471 &  4.1 & \bfseries 13.644 &  4.0 & \bfseries 13.641
& 1.319 & 1.80 & \bfseries 3.170 & 1.79 & \bfseries 3.170 \\

5-mix

& -7.713 & 0.3 & \bfseries 10.744 & 0.2 & \bfseries 10.735
& -6.651 & 1.9 & \bfseries 12.205 & 1.9 & \bfseries 12.201
& 1.309 & 1.64 & \bfseries 2.942 & 1.64 & \bfseries 2.942 \\

\bottomrule
\end{tabular*}

\caption{
Results on WSJ0-\{1,2,3,4,5\}mix-NR. Superscript $\star$ denotes using an estimated number of speakers for separation.
}
\label{tab:ss_results}
\end{table*}

\section{Evaluation Results}

This section reports the results of REnet, RDnet and REDnet.

\subsection{Results of RDnet}

Table \ref{tab:rd_ablation} reports the results of RDnet on WSJ0-\{2,3,4,5\}mix, assuming known speaker count at run time.
In row $1$, we replace the gating operation with an addition-based residual connection.
Its performance is consistently worse than row $2$, which turns on the gating.
Row $3$ and $4$ show that replacing PReLU with SiLU and introducing the Conv-SwiGLU block progressively improve separation.
Row $5-7$ investigate the contributions of different training objectives.
In row $5$, including the triple-path speaker interaction module by turning on the $\mathcal{L}_{\mathrm{TP}}$ loss and turning off $\mathcal{L}_{\mathrm{RD}}$ produces clear gains.
In row $6$, retaining a small weight for the intermediate $\mathcal{L}_{\mathrm{RD}}$ loss enhances overall performance.
In row $7$, further including $\mathcal{L}_{\mathrm{RM}}$ obtains clear improvement in the $4$- and $5$-mix conditions.
Finally, in row $8$, increasing the weight of $\mathcal{L}_{\mathrm{BCE}}$ yields the best overall results.
In addition, we observe that doing this can stabilize the training of speaker detection.

Table \ref{tab:wsj0_mix_sota} compares RDnet with existing methods on WSJ0-\{2,3,4,5\}mix, where all the systems need to do speaker counting.
RDnet obtains better SI-SDRi than the previous best in the $\{3,4,5\}$-mix cases, and a competitive SI-SDRi in the $2$-mix case.
The improvement is clear in the most challenging $5$-mix case (i.e., $20.8$ vs. $19.9$ dB SI-SDRi).
In addition, the speaker counting accuracy of RDnet is high, reaching $99.2\%$ in the $5$-mix case. 
This explains why RDnet has a minor performance drop when switching from oracle to estimated number of speakers (e.g., from $20.9$ to $20.8$ dB SI-SDRi in the $5$-mix case).
Table \ref{tab:computational_cost} reports the computation cost of RDnet.
It uses less computation but obtains better SI-SDRi than SR-CorrNet-B \cite{Shin2026SRCorrNet}, the previous best.

In Table \ref{tab:ss_results}, RDnet 
outperforms MUSE \cite{saijo2023single}
on WSJ0-\{1,2,3,4,5\}mix-NR, confirming its effectiveness and robustness in challenging noisy-reverberant conditions.

\begin{table*}[!]
\centering
\scriptsize
\sisetup{
    round-mode=places,
    round-precision=1,
    minimum-decimal-digits=1,
    list-separator={/},
    list-pair-separator={/},
    list-final-separator={/}
}
\setlength{\tabcolsep}{5pt}
\renewcommand{\arraystretch}{1.08}

\centering
\scriptsize
\setlength{\tabcolsep}{1.2pt}
\sisetup{
    mode=math,                    
    detect-weight=true,
    detect-inline-weight=math,
    round-mode=places,
    round-precision=1,
    minimum-decimal-digits=1,
    table-format=2.1,
    table-number-alignment=center,
    list-separator={/},
    list-pair-separator={/},
    list-final-separator={/}
}
\begin{tabular*}{\textwidth}{
    @{\extracolsep{\fill}}
    l
    c
    *{5}{
        S[
            table-format=-2.1,
            round-mode=places,
            round-precision=1,
            minimum-decimal-digits=1
        ]
        S[
            table-format=1.3,
            round-mode=places,
            round-precision=3, 
            minimum-decimal-digits=3
        ]
        S[
            table-format=1.2,
            round-mode=places,
            round-precision=2,
            minimum-decimal-digits=2
        ]
    }
    @{}
}

\toprule

& \multirow{2}{*}[-5.5pt]{\begin{tabular}[c]{@{}c@{}}Prior\\$C$?\end{tabular}}
& \multicolumn{3}{c}{{WHAM! (1--1)}}
& \multicolumn{3}{c}{{WHAMR!-A (1--2)}}
& \multicolumn{3}{c}{{WHAMR!-R (1--2)}}
& \multicolumn{3}{c}{{CHiME-4 (1--4)}}
& \multicolumn{3}{c}{{CHiME-4 (1--5)}} \\

\cmidrule(lr){3-5}
\cmidrule(lr){6-8}
\cmidrule(lr){9-11}
\cmidrule(lr){12-14}
\cmidrule(lr){15-17}

Method
& 
& {SDR} & {STOI} & {PESQ}
& {SDR} & {STOI} & {PESQ}
& {SDR} & {STOI} & {PESQ}
& {SDR} & {STOI} & {PESQ}
& {SDR} & {STOI} & {PESQ} \\

\midrule

Mixture
& --
& -0.91 & 0.761 & 1.11
& -0.92 & 0.760 & 1.11
& -0.04 & 0.731 & 1.11
& 7.55 & 0.870 & 1.27
& 7.55 & 0.870 & 1.27 \\

\midrule

USES
& \cmark
& 10.2 & 0.857 & 1.65
& 15.8 & 0.964 & 2.55
& 13.8 & 0.960 & 2.51
& 18.3 & 0.966 & 2.46
& 18.3 & 0.978 & 2.95 \\

TUSS
& \cmark
& 13.6 & 0.942 & 2.29
& 13.6 & 0.941 & 2.28
& 12.4 & 0.937 & 2.24
& 17.1 & 0.957 & 2.31
& 17.1 & 0.957 & 2.31 \\

FlexIO
& \cmark
& 13.8 & 0.944 & 2.31
& 15.8 & 0.960 & 2.61
& 14.9 & 0.958 & 2.58
& 21.3 & 0.980 & 2.86
& \bfseries 22.3 & 0.983 & 2.96 \\

\midrule

REDnet
& \cmark
& \bfseries 16.36 & \bfseries 0.965 & \bfseries 2.63
& \bfseries 18.07 & \bfseries 0.974 & \bfseries 2.91
& \bfseries 17.46 & \bfseries 0.973 & \bfseries 2.93
& \bfseries 21.36 & \bfseries 0.982 & \bfseries 3.21
& 22.22 & \bfseries 0.985 & \bfseries 3.30 \\

REDnet
& \xmark
& \bfseries 16.36 & \bfseries 0.965 & \bfseries 2.63
& \bfseries 18.07 & \bfseries 0.974 & \bfseries 2.91
& \bfseries 17.46 & \bfseries 0.973 & \bfseries 2.93
& \bfseries 21.36 & \bfseries 0.982 & \bfseries 3.21
& 22.22 & \bfseries 0.985 & \bfseries 3.30 \\

\bottomrule
\end{tabular*}

\caption{
SDR (dB), STOI and WB-PESQ results of REDnet on speech enhancement tasks (i.e., $C=1$). 
The numbers in parentheses beside each dataset denote the numbers of speakers and microphones, i.e., ``($C$--$P$)''.
}
\label{tab:red_speech_enhancement}

\vspace{0.3cm}

\centering
\scriptsize
\sisetup{
    mode=math,
    detect-weight=true,
    detect-inline-weight=math,
    round-mode=places,
    round-precision=1,
    minimum-decimal-digits=1,
    list-separator={/},
    list-pair-separator={/},
    list-final-separator={/}
}
\setlength{\tabcolsep}{1.0pt}
\begin{tabular*}{\textwidth}{
    @{\extracolsep{\fill}}
    l
    c
    *{5}{
        S[
            table-format=-2.1,
            round-mode=places,
            round-precision=1,
            minimum-decimal-digits=1
        ]
        S[
            table-format=-2.1,
            round-mode=places,
            round-precision=1,
            minimum-decimal-digits=1
        ]
        S[
            table-format=1.2,
            round-mode=places,
            round-precision=2,
            minimum-decimal-digits=2
        ]
    }
    @{}
}
\toprule

& \multirow{2}{*}[-5.5pt]{\begin{tabular}[c]{@{}c@{}}Prior\\$C$?\end{tabular}}

& \multicolumn{3}{c}{
    \shortstack{{WHAMR!-R} (2--1)}
}
& \multicolumn{3}{c}{
    \shortstack{{WHAMR!-R} (2--2)}
}
& \multicolumn{3}{c}{
    \shortstack{{WSJ1-CHiME} (2--2)}
}
& \multicolumn{3}{c}{
    \shortstack{{WSJ1-CHiME} (2--4)}
}
& \multicolumn{3}{c}{
    \shortstack{{WSJ1-CHiME} (3--3)}
} \\

\cmidrule(lr){3-5}
\cmidrule(lr){6-8}
\cmidrule(lr){9-11}
\cmidrule(lr){12-14}
\cmidrule(lr){15-17}

Method
&
& {SDR} & {SIR} & {PESQ}
& {SDR} & {SIR} & {PESQ}
& {SDR} & {SIR} & {PESQ}
& {SDR} & {SIR} & {PESQ}
& {SDR} & {SIR} & {PESQ} \\

\midrule

{Mixture} 
& {--}
& -4.05 & 0.08 & 1.08
& -4.05 & 0.08 & 1.08
& -1.26 & 0.05 & 1.17
& -1.17 & 0.05 & 1.17
& -4.22 & -3.24 & 1.12 \\

\midrule

{DNN-IVA}
& \cmark
& {--} & {--} & {--}
& {--} & {--} & {--}
& 10.7 & 24.1 & {--}
& {--} & {--} & {--}
& 7.7 & 20.1 & {--} \\

{TUSS}
& \cmark
& 9.5 & 25.3 & 1.82
& 9.5 & 25.3 & 1.82
& 14.9 & 29.1 & 2.73
& 15.1 & 29.4 & 2.75
& 11.4 & 23.1 & 2.15 \\

{FlexIO}
& \cmark
& 9.7 & 25.5 & 1.84
& 12.5 & 29.6 & 2.22
& \bfseries 19.6 & 35.0 & 3.36
& \bfseries 21.6 & 37.3 & 3.60
& 17.3 & 30.4 & 3.03 \\

\midrule

{REDnet}
& \bfseries \cmark
& \bfseries 13.80 & \bfseries 32.67 & \bfseries 2.34
& \bfseries 15.69 & \bfseries 34.41 & \bfseries 2.65
& \bfseries 19.58 & \bfseries 37.38 & \bfseries 3.49
& 21.15 & \bfseries 38.28 & \bfseries 3.69
& \bfseries 18.24 & \bfseries 33.99 & \bfseries 3.24 \\

{REDnet}
& \bfseries \xmark
& \bfseries 13.71 & \bfseries 32.59 & \bfseries 2.34
& \bfseries 15.65 & \bfseries 34.37 & \bfseries 2.65
& \bfseries 19.58 & \bfseries 37.38 & \bfseries 3.49
& 21.15 & \bfseries 38.28 & \bfseries 3.69
& \bfseries 18.24 & \bfseries 33.99 & \bfseries 3.24 \\
\bottomrule
\end{tabular*}

\caption{
SDR (dB), SIR (dB) and WB-PESQ results of REDnet on speech separation tasks (i.e., $C>1$).
}
\label{tab:red_speech_separation}
\end{table*}
\subsection{Results of REnet}

Table \ref{tab:re-fixed} reports the performance of REnet on the \textit{TAC-ad-hoc} dataset.
Compared with TAC under a similar computational cost, REnet consistently achieves better performance.
Table \ref{tab:re-fixed} also reports the performance on the \textit{TAC-fixed-array} dataset.
Although both REnet and TAC are trained exclusively on TAC-ad-hoc, REnet consistently outperforms TAC,
suggesting good generalizability to unseen array geometries.

\subsection{Results of REDnet}

Table \ref{tab:red_speech_enhancement} reports the results of REDnet for speech enhancement (i.e., $C=1$) on various datasets.
Even using an estimated number of speakers for inference, REDnet consistently outperforms state-of-the-art models including USES \cite{Zhang2023Diverse}, TUSS \cite{Saijo2025TUSS} and FlexIO \cite{masuyama2026flexio} (which assume a known speaker count) across most datasets and metrics.
Table \ref{tab:red_speech_separation} reports speech separation performance (i.e., $C>1$).
REDnet achieves, in most cases,
superior results over DNN-IVA \cite{Scheibler2021}, TUSS and FlexIO.
In addition, in both tables the performance of REDnet degrades only slightly when using estimated speaker count for inference.
These results show its effectiveness and robustness to variations in the number of microphones and speakers.
Table \ref{tab:chime4_dnsmos} reports DNSMOS scores on the simulated and real test sets of CHiME-4.
Although trained only on simulated data, REDnet obtains strong results on real recordings, outperforming USES, U2-C and FlexIO.

\begin{table}[t]
\centering
\scriptsize
\setlength{\tabcolsep}{2pt}
\sisetup{table-format=2.1,round-mode=places,round-precision=1,table-number-alignment=center,detect-weight=true,detect-inline-weight=math}

\begin{tabular*}{1\columnwidth}{
    @{\extracolsep{\fill}}
    l
    c
    S[table-format=1.2,round-precision=2]
    S[table-format=1.2,round-precision=2]
    S[table-format=1.2,round-precision=2]
    S[table-format=1.2,round-precision=2]
    @{}
}
\toprule
&
& \multicolumn{2}{c}{Simulated}
& \multicolumn{2}{c}{Real} \\
\cmidrule(lr){3-4}
\cmidrule(lr){5-6}
Method
& Prior $C$?
& {1ch} & {5ch} & {1ch} & {5ch} \\
\midrule

MVDR \cite{Masuyama2026}
& \cmark
& 2.08 & 2.57
& 1.46 & 1.95 \\

USES \cite{Zhang2023Diverse}
& \cmark
& 3.03 & 3.22
& 3.07 & 1.58 \\

U2-C \cite{Zhang2024U2C}
& \cmark
& {--} & {--}
& {--} & 3.08 \\

FlexIO \cite{masuyama2026flexio}
& \cmark
& 3.11 & 3.17
& 2.91 & 3.05 \\

\midrule
REDnet
& \xmark

& \bfseries 3.27 & \bfseries 3.29
& \bfseries 3.15 & \bfseries 3.10 \\

\bottomrule
\end{tabular*}

\caption{DNSMOS scores on CHiME-4 test sets.}
\label{tab:chime4_dnsmos}
\end{table}

\section{Conclusions}

We have proposed REDnet, a simple and effective model for speech separation under unknown numbers of speakers and variable numbers of microphones.
State-of-the-art separation performance is obtained on multiple representative datasets in speech separation and enhancement.
Future research will focus on
better exploiting spatial cues in RE and extending RD for unsupervised speech separation.

\bibliography{aaai2027}

@string{icassp = "Proc. ICASSP"}

@string{interspeech = "Proc. Interspeech"}

@string{iwaenc = "Proc. IWAENC"}

@string{asru = "Proc. ASRU"}

@string{ieee-taslp = "IEEE Trans. Audio, Speech, Lang. Process."}

@string{ieee-acm-taslp = "IEEE/ACM Trans. Audio, Speech, Lang. Process."}

@string{ieee-spl = "IEEE Signal Process. Lett."}

@string{icml = "Proc. ICML"}

@string{iclr = "Proc. ICLR"}

@string{nips = "Proc. NIPS"}

@string{aaai = "Proc. AAAI"}

@string{csl = "Comput. Speech Lang."}

@string{slt = "Proc. SLT"}

@string{ijcai = "Proc. IJCAI"}

@string{eusipco = "Proc. EUSIPCO"}

@article{McDermott2009,
author = {{McDermott}, Josh},
issn = {08997667},
journal = {Current Biology},
number = {22},
pages = {1024--1027},
pmid = {15992485},
title = {{The Cocktail Party Problem}},
volume = {19},
year = {2009}
}

@inproceedings{Yoshioka2022VarArray,
author = {Yoshioka, Takuya and Wang, Xiaofei and Wang, Dongmei and Tang, Min and Zhu, Zirun and Chen, Zhuo and Kanda, Naoyuki},
title = {{VarArray: Array-Geometry-Agnostic Continuous Speech Separation}},
booktitle = icassp,
pages = {6027--6031},
year = {2022},
}

@inproceedings{lee2024boosting,
title={{Boosting Unknown-Number Speaker Separation with Transformer Decoder-Based Attractor}},
author={Lee, Younglo and Choi, Shukjae and Kim, Byeong-Yeol and Wang, Zhong-Qiu and Watanabe, Shinji},
booktitle= icassp,
pages={446--450},
year={2024},
}

@inproceedings{masuyama2026flexio,
title={{FlexIO: Flexible Single- and Multi-Channel Speech Separation and Enhancement}},
author={Masuyama, Yoshiki and Saijo, Kohei and Paissan, Francesco and Han, Jiangyu and Delcroix, Marc and Aihara, Ryo and Germain, Fran{\c{c}}ois G and Wichern, Gordon and Le Roux, Jonathan},
booktitle= icassp,
pages={14417--14421},
year={2026},
}

@article{Barker2017CHiME3,
author = {Barker, Jon and Marxer, Ricard and Vincent, Emmanuel and Watanabe, Shinji},
journal = csl,
pages = {605--626},
title = {{The Third ‘CHiME' Speech Separation and Recognition Challenge: Analysis and Outcomes}},
volume = {46},
year = {2017}
}

@inproceedings{Scheibler2021,
author = {Scheibler, Robin and others},
booktitle = icassp,
pages = {176--180},
title = {{Surrogate Source Model Learning for Determined Source Separation}},
year = {2021}
}

@inproceedings{yang25SepNetEDCI,
  title     = {{Speaker Separation for an Unknown Number of Speakers with Encoder-Decoder-Based Contextual Information Module}},
  author    = {Xue Yang and others},
  year      = {2025},
  booktitle = interspeech,
  pages     = {1448--1452},
}

@inproceedings{rix2001NBPESQ,
  title={{Perceptual Evaluation of Speech Auality (PESQ)-A New Method for Speech Quality Assessment of Telephone Networks and Codecs}},
  author={Rix, Antony W and Beerends, John G and Hollier, Michael P and Hekstra, Andries P},
  booktitle= icassp,
  pages={749--752},
  year={2001},
}

@article{taal2011STOI,
  title={{An Algorithm for Intelligibility Prediction of Time--Frequency Weighted Noisy Speech}},
  author={Taal, Cees H and Hendriks, Richard C and Heusdens, Richard and Jensen, Jesper},
  journal= ieee-taslp,
  volume={19},
  number={7},
  pages={2125--2136},
  year={2011},
}

@inproceedings{reddy2022dnsmos,
  title={{DNSMOS P. 835: A Non-Intrusive Perceptual Objective Speech Quality Metric to Evaluate Noise Suppressors}},
  author={Reddy, Chandan KA and others},
  booktitle= icassp,
  pages={886--890},
  year={2022},
}

@techreport{ITU2007WBPESQ,
  author      = {{ITU}},
  title       = {Wideband Extension to Recommendation {P.862} for the
                 Assessment of Wideband Telephone Networks and Speech Codecs},
  institution = {International Telecommunication Union},
  type        = {{ITU-T} Recommendation},
  number      = {P.862.2},
  address     = {Geneva, Switzerland},
  year        = {2007}
}

@inproceedings{kim2023investigation,
  title={{Investigation of Training Mute-Expressive End-to-End Speech Separation Networks for an Unknown Number of Speakers}},
  author={Kim, Younggwan and others},
  booktitle= interspeech,
  pages={3764--3768},
  year={2023}
}

@inproceedings{luo2020separating,
  title={{Separating Varying Numbers of Sources with Auxiliary Autoencoding Loss}},
  author={Luo, Yi and others},
  booktitle = interspeech,
  pages = {2622--2626},
  year={2020}
}

@misc{Shin2026SRCorrNet,
title={{Asymmetric Encoder-Decoder Based on Time-Frequency Correlation for Speech Separation}},
author={Shin, Ui-Hyeop and others},
year={2026},
eprint={2603.29097},
archivePrefix={arXiv},
primaryClass={eess.AS},
url={https://arxiv.org/abs/2603.29097},
}

@misc{jin2022coarse,
archivePrefix = {arXiv},
arxivId = {2203.16054},
author = {Jin, Zhenhao and others},
booktitle = {arXiv preprint arXiv:2203.16054},
eprint = {2203.16054},
title = {{Coarse-to-Fine Recursive Speech Separation for Unknown Number of Speakers}},
year = {2022}
}

@inproceedings{Heymann2016smeab,
author = {Heymann, Jahn and others},
    title = {{Neural Network based Spectral Mask Estimation for Acoustic Beamforming}},
    booktitle = icassp,
    pages = {196--200},
    year = {2016}
}

@article{Wang2018iCombineSpectralSpatial,
author = {Wang, Zhong-Qiu and Wang, DeLiang},
journal = ieee-acm-taslp,
number = {2},
pages = {457--468},
title = {{Combining Spectral and Spatial Features for Deep Learning Based Blind Speaker Separation}},
volume = {27},
year = {2019}
}

@inproceedings{Erodgan2016MVDR,
author = {Erdogan, Hakan and Hershey, John R. and Watanabe, Shinji and Mandel, Michael I. and {Le Roux}, Jonathan},
title = {{Improved MVDR Beamforming using Single-Channel Mask Prediction Networks}},
booktitle = interspeech,
pages = {1981--1985},
year = {2016}
}

@inproceedings{shi2018listen,
title={{Listen, Think and Listen Again: Capturing Top-Down Auditory Attention for Speaker-Independent Speech Separation.}},
author={Shi, Jing and Xu, Jiaming and Liu, Guangcan and Xu, Bo and others},
booktitle= ijcai,
pages={4353--4360},
year={2018}
}

@article{Masuyama2026,
author = {Masuyama, Yoshiki and Chang, Xuankai and Zhang, Wangyou and Cornell, Samuele and Wang, Zhong-Qiu and Ono, Nobutaka and Qian, Yanmin and Watanabe, Shinji},
issn = {10958363},
journal = csl,
number = {101813},
pages = {1--18},
title = {{An End-to-End Integration of Speech Separation and Recognition with Self-Supervised Learning Representation}},
volume = {95},
year = {2026}
}

@article{chetupalli2023speaker,
  title={{Speaker Counting and Separation from Single-Channel Noisy Mixtures}},
  author={Chetupalli, Srikanth Raj and others},
  journal= ieee-acm-taslp,
  volume={31},
  pages={1681--1692},
  year={2023},
}

@inproceedings{alizadeh2025resepnet,
  title={{ReSepNet: A Unified-Light Model for Recursive Speech Separation with Unknown Speaker Count}},
  author={Alizadeh, Hadi and others},
  booktitle= interspeech,
  pages={1458--1462},
  year={2025}
}

@inproceedings{kinoshita2018listening,
title={{Listening to Each Speaker One by One with Recurrent Selective Hearing Networks}},
author={Kinoshita, Keisuke and Drude, Lukas and Delcroix, Marc and Nakatani, Tomohiro},
booktitle= icassp,
pages={5064--5068},
year={2018},
}

@inproceedings{Nachmani2020gatedDPRNN,
author = {Nachmani, Eliya and others},
booktitle = icml,
pages = {7121--7132},
title = {{Voice Separation with an Unknown Number of Multiple Speakers}},
year = {2020}
}

@inproceedings{Maciejewski2020,
archivePrefix = {arXiv},
arxivId = {1910.10279},
author = {MacIejewski, Matthew and others},
booktitle = icassp,
doi = {10.1109/ICASSP40776.2020.9053327},
eprint = {1910.10279},
issn = {15206149},
pages = {696--700},
title = {{WHAMR!: Noisy and Reverberant Single-Channel Speech Separation}},
year = {2020}
}

@article{Kolbak2017,
archivePrefix = {arXiv},
arxivId = {1703.06284},
author = {Kolb{\ae}k, Morten and Yu, Dong and Tan, Zheng-Hua and Jensen, Jesper},
doi = {10.1109/TASLP.2017.2726762},
eprint = {1703.06284},
issn = {23299290},
journal = ieee-acm-taslp,
number = {10},
pages = {1901--1913},
title = {{Multitalker Speech Separation with Utterance-Level Permutation Invariant Training of Deep Recurrent Neural Networks}},
volume = {25},
year = {2017}
}

@inproceedings{Wichern2019,
author = {Wichern, Gordon and Antognini, Joe and Flynn, Michael and Zhu, Licheng Richard and McQuinn, Emmett and Crow, Dwight and Manilow, Ethan and {Le Roux}, Jonathan},
booktitle = interspeech,
pages = {1368--1372},
title = {{WHAM!: Extending Speech Separation to Noisy Environments}},
year = {2019}
}

@inproceedings{takahashi2019recursive,
author = {Takahashi, Naoya and Parthasaarathy, Sudarsanam and Goswami, Nabarun and Mitsufuji, Yuki},
booktitle = interspeech,
pages = {1348--1352},
title = {{Recursive Speech Separation for Unknown Number of Speakers}},
year = {2019}
}

@inproceedings{Hershey2016,
author = {Hershey, John R. and Chen, Zhuo and {Le Roux}, Jonathan and Watanabe, Shinji},
booktitle = icassp,
pages = {31--35},
title = {{Deep Clustering: Discriminative Embeddings for Segmentation and Separation}},
year = {2016}
}

@inproceedings{zhu2021multi,
title={{Multi-Decoder DPRNN: Source Separation for Variable Number of Speakers}},
author={Zhu, Junzhe and others},
booktitle= icassp,
pages={3420--3424},
year={2021},
}

@inproceedings{chazan2021single,
title={{Single Channel Voice Separation for Unknown Number of Speakers under Reverberant and Noisy Settings}},
author={Chazan, Shlomo E and Wolf, Lior and Nachmani, Eliya and Adi, Yossi},
booktitle= icassp,
pages={3730--3734},
year={2021},
}

@article{kim2023training,
title={{On Training Speech Separation Models with Various Numbers of Speakers}},
author={Kim, Hyeonseung and Shin, Jong Won},
journal=ieee-spl,
volume={30},
pages={1202--1206},
year={2023},
}

@inproceedings{Pandey2022TADRN,
title={{Time-Domain Ad-Hoc Array Speech Enhancement Using a Triple-Path Network}},
author={Pandey, Ashutosh and Xu, Buye and Kumar, Anurag and Donley, Jacob and Calamia, Paul and Wang, DeLiang},
booktitle= interspeech,
pages={729--733},
year={2022}
}

@inproceedings{wang2020neural,
title={{Neural Speech Separation using Spatially Distributed Microphones}},
author={Wang, Dongmei and others},
booktitle= interspeech,
pages={339--343},
year={2020}
}

@inproceedings{wang23SDNet,
title={{SDNet: Stream-Attention and Dual-Feature Learning Network for Ad-Hoc Array Speech Separation}},
author={Honglong Wang and Chengyun Deng and Yanjie Fu and Meng Ge and Longbiao Wang and Gaoyan Zhang and Jianwu Dang and Fei Wang},
year={2023},
booktitle= interspeech,
pages={3739--3743},
}

@inproceedings{guo2024graph,
title={Graph Attention Based Multi-Channel U-Net for Speech Dereverberation With Ad-Hoc Microphone Arrays.},
author={Guo, Hongmei and Chen, Yijiang and Zhang, Xiaolei and Li, Xuelong},
booktitle= interspeech,
pages={617--621},
year={2024}
}

@inproceedings{guo2025co,
title={Co-Attention Based Multi-Channel TF-GridNet for Speech Separation with Ad-Hoc Microphone Arrays},
author={Guo, Hongmei and Feng, Linfeng and Chen, Yijiang and Li, Xueqing and Zhu, Boyu and Wang, Hao-Yu and Zhang, Xiao-Lei and Li, Xuelong},
booktitle= icassp,
pages={1--5},
year={2025},
}

@article{WDLreview,
archivePrefix = {arXiv},
arxivId = {1708.07524},
author = {Wang, DeLiang and Chen, Jitong},
doi = {10.1109/TASLP.2018.2842159},
eprint = {1708.07524},
issn = {23299290},
journal = ieee-acm-taslp,
number = {10},
pages = {1702--1726},
title = {{Supervised Speech Separation Based on Deep Learning: An Overview}},
volume = {26},
year = {2018}
}

@inproceedings{Li2026sepprune,
author = {Li, Yuqi and Li, Kai and Yin, Xin and Yang, Zhifei and Dong, Zeyu and Yao, Zhengtao and Xu, Haoyan and Tian, Yingli and Lu, Yao},
title = {{SepPrune: Structured Pruning for Efficient Deep Speech Separation}},
year = {2026},
booktitle = aaai,
pages = {31861--31869}
}

@inproceedings{li2023pgss,
  title={{Pgss: Pitch-Guided Speech Separation}},
  author={Li, Xiang and Wang, Yiwen and Sun, Yifan and Wu, Xihong and Chen, Jing},
  booktitle= aaai,
  pages={13130--13138},
  year={2023}
}

@inproceedings{fan2025bsdb,
  title={{BSDB-Net: Band-Split Dual-Branch Network with Selective State Spaces Mechanism for Monaural Speech Enhancement}},
  author={Fan, Cunhang and Liu, Enrui and Li, Andong and Tao, Jianhua and Zhou, Jian and Li, Jiahao and Zheng, Chengshi and Lv, Zhao},
  booktitle= aaai,
  pages={23850--23858},
  year={2025}
}

@inproceedings{xu2025arraydps,
  title={ArrayDPS: Unsupervised Blind Speech Separation with a Diffusion Prior},
  author={Xu, Zhongweiyang and Fan, Xulin and Wang, Zhong Qiu and Jiang, Xilin and Choudhury, Romit Roy},
  booktitle= icml,
  pages={69160--69188},
  year={2025},
}

@inproceedings{xu2025tiger,
  title={Tiger: Time-frequency Interleaved Gain Extraction and Reconstruction for Efficient Speech Separation},
  author={Xu, Mohan and Li, Kai and Chen, Guo and Hu, Xiaolin},
  booktitle= iclr,
  pages={70205--70222},
  year={2025}
}

@inproceedings{LeRoux2019,
author = {{Le Roux}, Jonathan and Wisdom, Scott and Erdogan, Hakan and {Hershey}, John R},
booktitle = icassp,
pages = {626--630},
title = {{SDR - Half-Baked or Well Done?}},
year = {2019}
}

@inproceedings{Luo2019FasNet,
author = {Luo, Yi and Ceolini, Enea and Han, Cong and Liu, Shih-Chii and Mesgarani, Nima},
title = {{FaSNet: Low-Latency Adaptive Beamforming for Multi-Microphone Audio Processing}},
booktitle = asru,
pages = {260--267},
year = {2019},
}

@inproceedings{Zhang2023Diverse,
author = {Zhang, Wangyou and Saijo, Kohei and Wang, Zhong-Qiu and Watanabe, Shinji and Qian, Yanmin},
title = {{Toward Universal Speech Enhancement for Diverse Input Conditions}},
booktitle = asru,
pages = {1--6},
year = {2023},
}

@inproceedings{Saijo2025TUSS,
author = {Saijo, Kohei and Ebbers, Janek and Germain, Fran{\c{c}}ois G. and Wichern, Gordon and {Le Roux}, Jonathan},
booktitle = icassp,
title = {{Task-Aware Unified Source Separation}},
year = {2025}
}

@inproceedings{horiguchi2020end,
title={{End-to-End Speaker Diarization for an Unknown Number of Speakers with Encoder-decoder Based Attractors}},
author={Horiguchi, Shota and Fujita, Yusuke and Watanabe, Shinji and Xue, Yawen and Nagamatsu, Kenji},
booktitle= interspeech,
pages={269--273},
year={2020}
}

@inproceedings{Luo2020e2e,
archivePrefix = {arXiv},
arxivId = {1910.14104},
author = {Luo, Yi and Chen, Zhuo and Mesgarani, Nima and Yoshioka, Takuya},
booktitle = icassp,
doi = {10.1109/icassp40776.2020.9054177},
eprint = {1910.14104},
pages = {6394--6398},
title = {{End-to-End Microphone Permutation and Number Invariant Multi-Channel Speech Separation}},
year = {2020}
}

@inproceedings{Zhang2024U2C,
author = {Zhang, Wangyou and others},
booktitle = icassp,
pages = {10696--10700},
title = {{Improving Design of Input Condition Invariant Speech Enhancement}},
year = {2024}
}

@article{vincent2006bssevalsdr,
  title={{Performance Measurement in Blind Audio Source Separation}},
  author={Vincent, Emmanuel and others},
  journal= ieee-taslp,
  volume={14},
  number={4},
  pages={1462--1469},
  year={2006},
}

@article{Wang2023TFGridNet,
author = {Wang, Zhong-Qiu and Cornell, Samuele and Choi, Shukjae and Lee, Younglo and Kim, Byeong-Yeol and Watanabe, Shinji},
journal = ieee-acm-taslp,
pages = {3221--3236},
title = {{TF-GridNet: Integrating Full- and Sub-Band Modeling for Speech Separation}},
volume = {31},
year = {2023}
}

@inproceedings{saijo2023single,
title={{A Single Speech Enhancement Model Unifying Dereverberation, Denoising, Speaker Counting, Separation, and Extraction}},
author={Saijo, Kohei and Zhang, Wangyou and Wang, Zhong-Qiu and Watanabe, Shinji and Kobayashi, Tetsunori and Ogawa, Tetsuji},
booktitle= asru,
pages={1--6},
year={2023},
}

@inproceedings{Saijo2024LocoFormer,
author = {Saijo, Kohei and Wichern, Gordon and Germain, Fran{\c{c}}ois G. and Pan, Zexu and Roux, Jonathan Le},
booktitle = iwaenc,
pages = {205--209},
title = {{TF-Locoformer: Transformer with Local Modeling by Convolution for Speech Separation and Enhancement}},
year = {2024}
}

@inproceedings{kim2024enhanced,
title={Enhanced Deep Speech Separation in Clustered Ad Hoc Distributed Microphone Environments},
author={Kim, Jihyun and Kindt, Stijn and Madhu, Nilesh
      and Kang, Hong-Goo},
booktitle= interspeech,
pages={2185--2189},
year={2024},
}

@article{ELFWING2018silu,
title = {{Sigmoid-Weighted Linear Units for Neural Network Function Approximation in Reinforcement Learning}},
journal ={Neural Networks},
volume = {107},
pages = {3-11},
year = {2018},
author = {Stefan Elfwing and others},
}

@inproceedings{shi2020sequence,
title={{Sequence to Multi-sequence Learning via Conditional Chain Mapping for Mixture Signals}},
author={Shi, Jing and Chang, Xuankai and Guo, Pengcheng and Watanabe, Shinji and Fujita, Yusuke and Xu, Jiaming and Xu, Bo and Xie, Lei},
booktitle= nips,
volume={33},
pages={3735--3747},
year={2020}
}

@inproceedings{Maiti2023eendss,
author={Maiti, Soumi and Ueda, Yushi and Watanabe, Shinji and Zhang, Chunlei and Yu, Meng and Zhang, Shi-Xiong and Xu, Yong},
booktitle= slt, 
title={{EEND-SS: Joint End-to-End Neural Speaker Diarization and Speech Separation for Flexible Number of Speakers}}, 
year={2023},
pages={480-487},
}

@inproceedings{chetupalli2022sepeda,
title={{Speech Separation for an Unknown Number of Speakers using Transformers With Encoder-Decoder Attractors.}},
author={Chetupalli, Srikanth Raj and others},
booktitle= interspeech,
pages={5393--5397},
year={2022}
}

@inproceedings{wang2021continuous,
  title={{Continuous Speech Separation with Ad Hoc Microphone Arrays}},
  author={Wang, Dongmei and Yoshioka, Takuya and Chen, Zhuo and Wang, Xiaofei and Zhou, Tianyan and Meng, Zhong},
  booktitle=eusipco,
  pages={1100--1104},
  year={2021},
}

\end{document}